\documentstyle[prl,aps,12pt]{revtex}
\begin{document}      

\draft
\vspace{2cm}
\title{On the Renormalization of a Bosonized Version of the\\
 Chiral Fermion-Meson Model at Finite Temperature}
\vspace{4cm}
\author{\it H.C.G. Caldas$^{(a)}$\thanks{e-mail: hcaldas@funrei.br} and M.C. Nemes$^{(b)}$}

\address{${(a)}$Departamento de Ci\^{e}ncias Naturais, DCNAT \\
Funda\c{c}\~ao de Ensino Superior de S\~{a}o Jo\~{a}o del Rei, FUNREI,\\ 
Pra\c{c}a Dom Helv\'ecio, 74, CEP:36300-000, S\~{a}o Jo\~{a}o del Rei, MG, Brazil}

\address{${(b)}$Departamento de F\'\i sica, Universidade Federal de Minas Gerais, UFMG,\\
CP 702,CEP:30.161-970,Belo Horizonte, MG, Brazil}

\vspace{1cm}
\date{July, 2001}

\maketitle                          
\vspace*{0.5in}
\centerline{\bf Abstract}

\vspace{1cm}

Feynman's functional formulation of statistical mechanics
is used to study the renormalizability of the well known Linear Chiral Sigma Model in 
the presence of fermionic fields at finite temperature in an alternative way. It is shown that the renormalization conditions coincide with those of the zero temperature model.

\vspace{2cm}
\noindent
PACS numbers: 11.10.Wx, 12.39.Fe, 11.10.Gh

\newpage


\medskip

The advent of ultrarelativistic heavy ion collisions has given to finite temperature 
field theory (FTFT) a special task, since this kind of experiment may reveal a new state of 
matter which is the quark-gluon plasma. The question of what happens with the properties of the particles at non-zero temperature (immersed in a thermal medium) is one of the most important topics of FTFT. There have been important contributions in the area, mostly contend a field theoretical questions such as the derivation of particle properties. Some examples of this research subject through the last three decades can be found in \cite{Dolan,Weinberg,Bernard,Cornwall,Baym,Linde,Kapusta,Braaten0,Bellac,Das}. On the other hand, the inclusion of temperature in effective models has gained importance not only in the context of hadron physics, but also in cosmology and astroparticle physics for which a knowledge of the behavior of hadronic masses with temperature are most relevant.

Derivations of the pion and sigma meson masses as well as the condensate as a function of temperature have been given both from a phenomenological \cite{Caldas} as well as a theoretical point of view \cite{Caldas2}. An immediate question arises as to the eventual modifications introduced by the boson coupling to fermionic degrees of freedom. The first issue of concern about the inclusion of fermions is the presence of infinities which
will arise. How do they appear in the Feynman's functional formulations
of statistical mechanics? It is known that the chiral fermion meson 
model \cite{Gell-Mann} is renormalizable at zero \cite{Lee,Franco} and finite temperature \cite{Ram-Mohan,Caldas2}. In the early study of the renormalizability of the linear sigma model at finite temperature presented in ref \cite{Ram-Mohan}, it is 
shown that the renormalization conditions are the same as for the zero temperature case. We will show here in this reinvestigation of the well studied chiral fermion-meson model that this is also the case for its `` bosonized '' version. It is often emphasized that it is enough to renormalize the model at zero temperature since finite temperature does not introduce any new UV divergences to the theory. Namely, the zero temperature counter terms i.e., with the vacuum parameters, are sufficient to take care of the divergences at finite temperature as well.  However, it is well known that the perturbative expansion breaks down at finite temperature. 
That is, in higher order diagrams (where powers of the temperature over the intrinsic mass, $T/m$, increase making these diagrams larger than the lower ones at 
high T) or in theories with spontaneous symmetry breaking, there is the necessity of a resummation or, in other words, a sum of an infinite subset 
of diagrams. The reliability of perturbation theory can be recovered with such a resummation.  
In these `` resumed '' theories the counter terms need also to suffer the resummation if one wants achieve renormalization satisfactorily. This is because the mass which multiplies the divergence will acquire a temperature dependence through the gap equation for the mass. When the resummation is done consistently, the theory can be renormalized since the resumed temperature dependent counter terms have the same structure as in the ordinary perturbation theory \cite{Caldas2,Caldas3}. However, in ordinary perturbation theory at finite temperature to leading order, renormalization can be employed in the conventional fashion, as is the case in this work.

\vspace{1cm}
The Lagrangian density of the chiral linear sigma model is given by

\begin{eqnarray}
{\cal L }=\bar{\psi}(i\partial_{\mu} \gamma^{\mu} -g\Gamma \varphi )\psi +
 \frac{1}{2}(\partial_{\mu}\varphi)(\partial^{\mu}\varphi)-\frac{\lambda}{4}
(\varphi^{2}-f_{\pi}^{2})^{2},
\end{eqnarray}
 
\noindent where $\varphi =(\sigma,\vec{\pi}) $ and 
$\Gamma=(1,i\gamma_{5} \vec{\tau})$ contain the chiral mesonic fields
and their coupling to the fermionic fields respectively, $\lambda$ and 
$g$ are positive coupling constants and $f_\pi$ is the pion decay constant 
in vacuum. Next we notice that a convenient way to write the Lagrangian 
is in terms of an hermitean operator: 

\begin{eqnarray}
h \equiv \frac{\vec{\alpha }.\vec{\nabla }}{i}+g~\tilde \beta ~\Gamma ~\varphi,
\end{eqnarray}

\noindent where $\tilde \beta$ and $\vec{\alpha }$ are defined in terms of the usual Dirac
Matrices $\gamma^{\mu}=(\tilde \beta,\vec \gamma)$, such that:

\begin{eqnarray}
\tilde \beta = \gamma^0,\\
\nonumber
\vec{\alpha }= \tilde \beta \vec \gamma.
\end{eqnarray}
Then, written in terms of $h$ and $\tau = i t$ the Lagrangian becomes

\begin{eqnarray}
{\cal L } \to \psi^{\dagger}(\partial_{\tau } +h )\psi +
 \frac{1}{2}(\partial_{\mu }\varphi )^{2}+\frac{\lambda}{4}
(\varphi^{2}-f_{\pi }^{2})^{2}  \equiv  {\cal L}_E,
\end{eqnarray}
 
\noindent where $i\partial_{t}=-\partial_{\tau}$ and ${\cal L}_E$ is the Euclidean Lagrangian. 
We note that in this Euclidean metric, we have $g^{\mu \nu}=g_{\mu \nu}=I$ and consequently $x^{\mu}=x_{\mu}=(\tau,\vec r)$ and $\gamma^{\mu}=\gamma_{\mu}=(i \tilde \beta,\vec \gamma)$. As is well known, given the Lagrangian of a system with Euclidean metric
the quantum partition function is given by

\begin{eqnarray}
{\cal Z}=Tr~e^{-\beta H}=\int {\cal D}(\psi^{*}){\cal D}(\psi ){\cal D}(\varphi )
e^{iS(\psi^{\dagger },\psi,\varphi )}, 
\label{fupar}
\end{eqnarray}

\noindent where $\beta$ in (\ref{fupar}) is the inverse of the temperature $T$ and

\begin{eqnarray}
iS=-\int_{0}^{\beta } d\tau \int d^{3}r~{\cal L}_E.
\end{eqnarray}

We can immediately integrate out the fermionic fields

\begin{eqnarray}
\int{\cal D}(\psi^{*}){\cal D}(\psi )&~& exp~{ -\int_{0}^{\beta } 
d\tau \int d^{3}r
~\psi^{\dagger}(\partial_{\tau}+h)\psi }\nonumber\\
~&=& det(\partial_{\tau}+h)=exp~Tr~
\log (\partial_{\tau}+h).
\end{eqnarray}

The meaning of the trace operation should be well understood; to the
fermions in the discretized space-time are attached two indices $i$ and
$s$, $\psi_{is}=\psi_{s}({r_{i}},\tau_{i})$. The index $s$ stands for
all four Dirac indices, the isospin indices etc. In this case we define the
matrix $(\partial_{\tau}+h)_{is,js^{'}}$ through the expression

\begin{eqnarray}
\int_0 ^{\beta} d\tau \int d^{3}r~
\psi^{\dagger}(\partial_{\tau}+h)\psi =\sum_{is,js^{'}}
\psi^{\dagger}_{is}~(\partial_{\tau}+h)_{is,js^{'}}~\psi_{js^{'}},
\end{eqnarray}
so that we have the equivalence among the three expressions

\begin{eqnarray}
Tr~\log (\partial_{\tau}+h)&=&\sum_{is}~[\log ~(\partial_{\tau}+h)]_{is,is}
\nonumber\\
&=&\int^{\beta}_{0} d\tau \int d^{3}r \sum_{s}~\langle \vec{r}\tau |
~[\log (\partial
_{\tau} +h)]_{ss} ~|\vec{r}\tau \rangle \nonumber\\
~&=& \sum_{\vec{k},\omega,s}~\langle \vec{k}\omega |
~[\log (\partial_{\tau} +h)]_{ss}~|
\vec{k}\omega \rangle,
\end{eqnarray}

\noindent with

\begin{eqnarray}
\langle\vec{r}|\vec{k}\rangle =\frac{1}{\sqrt{\Omega}}~e^{i\vec{k}.\vec{r}},
~~~~~~~\langle\tau |\omega \rangle =\frac{1}{\sqrt{\beta}}~e^{-\omega \tau},
\end{eqnarray}
where $\vec{k}$ are discrete periodic momenta and $\omega $ the usual Matsubara
frequencies. So $Tr~\log (\partial_{\tau} +h)$ is a trace of the operator
$\partial_{\tau} +h$ in the one fermion state space where the operator
in question acts.

The integration over the fermions yields an effective action
$iS_{eff}(\varphi )$ which will depend only on the chiral fields
$\varphi $.

\begin{eqnarray}
{\cal Z}=\int {\cal D}[\varphi ]e^{iS_{eff}(\varphi )},
\end{eqnarray}
where

\begin{eqnarray}
iS_{eff}(\varphi )= Tr ~ \log (\partial_{\tau} +h ) ~
- \int^{\beta}_{0} d\tau \int d^{3}r~{\cal L}(\varphi)=\\
\nonumber 
-\int^{\beta}_{0} d\tau \int d^{3}r~
\left\{-tr~[\log (\partial_{\tau} +h )
] +\frac{1}{2}(\partial_{\mu}\varphi )^2 +
\frac{\lambda}{4}(\varphi^2 -f_{\pi}^2)^2 \right\} \equiv - \int d^4r {\cal L}_{eff}(\varphi),
\end{eqnarray}
where in the above expression we have defined $tr~log(\partial_{\tau} +h )=\sum_{s} ~log(\partial_{\tau} +h)_{ss}$, $\int d^4r=\int^{\beta}_{0} d\tau \int d^{3}r$ and

\begin{equation}
{\cal L}_{eff}(\varphi)=-tr~[\log (\partial_{\tau} +h )
] +\frac{1}{2}(\partial_{\mu}\varphi )^2 +
\frac{\lambda}{4}(\varphi^2 -f_{\pi}^2)^2. 
\end{equation}
The trace is over Dirac and isospin matrices. An explicit calculation of
the term $tr~\log (\partial_{\tau}+h)$ reveals the presence of infinities. 
We proceed thus to renormalization for that purpose. We develop the
effective action around the point we call the physical vacuum:

\begin{eqnarray}
\varphi_{0}=(f_{\pi} ,0,0,0).
\end{eqnarray}

Since $detA=det \widetilde A$, where $A$ is a given operator, with $\widetilde A = C~A~C^{-1}= \pm A$ ($C=i\gamma^2 \tilde \beta$ is the charge conjugation matrix in the Dirac representation \cite{Dirac}), we have
\begin{eqnarray}
tr ~ \log (\partial_\tau + h)= tr ~ \log (-\partial_\tau + h)= \frac{1}{2} 
tr ~ \log (-\partial_\tau + h)(\partial_\tau + h)=\\
\nonumber
\frac{1}{2} tr ~ \log (-\partial_{\mu}^{2}+g^2 \varphi^{2} + ig \gamma^{\mu} (\partial_{\mu}\varphi)\Gamma)
\end{eqnarray}
Here we write 

\begin{eqnarray}
&-&\partial_{\mu}^{2}+g^2 \varphi^{2} + ig \gamma^{\mu} (\partial_{\mu}\varphi
)\Gamma \nonumber\\
~&=& -\partial_{\mu}^{2}+g^2 \varphi_{0}^{2} + g^2 (\varphi^2 -
\varphi_{0}^2 ) +ig \gamma^{\mu}(\partial_{\mu}\varphi
)\Gamma \nonumber\\
~&=& G_{0}^{-1} +V,
\end{eqnarray}
with the ``unperturbed propagator'' defined as 

\begin{eqnarray}
G_{0}^{-1}=-\partial_{\mu}^{2}+g^2 \varphi_{0}^{2}
=-\partial_{\mu}^{2}+g^2 f_{\pi}^2,
\end{eqnarray}
and the ``perturbation''

\begin{eqnarray}
V &=& g^2 (\varphi^2 -\varphi_{0}^2 )+ ig \gamma^{\mu}
(\partial_{\mu}\varphi) \Gamma \nonumber\\
~&=&
g^2 (\sigma^2 +\vec{\pi}^2 -f_{\pi}^2 ) +ig
\gamma^{\mu}(\partial_{\mu}\sigma )-g
\gamma_{5}\gamma^{\mu}(\partial_{\mu}\vec{\pi})\cdot\vec{\tau}. 
\end{eqnarray}

Note that $G_{0}$, the unperturbed or the free particle propagator, which acts in the one 
fermion state space has the form of a boson propagator with a $gf_{\pi}$ mass. It is also 
important to remark that the perturbation $V$ is not a power of the coupling constant $g$, 
but proportional to the amplitude $\tilde{\varphi }=\varphi - \varphi_{0}
$, the deviation of the field with respect to its vacuum value. 
We can thus interpret what has been done as a bosonization in the fermionic one loop
approximation. The propagator $G_{0}$ is diagonal in the plane wave basis we have defined

\begin{eqnarray}
G_{0}|\vec{k}\omega s\rangle =\frac{1}{k^2 +g^2 f_{\pi}^2 -\omega^2
}~|\vec{k}\omega s\rangle =\frac{1}{e_{k}^2 -\omega^2 }~|\vec{k}\omega
s\rangle,
\end{eqnarray}
with $e_k =\sqrt{k^2 +g^2 f_{\pi}^2 }$.

\vspace{0.2cm}

The Lagrangian related to the effective action, eq.(12), can be
rewritten as 

\begin{eqnarray}
{\cal L}_{eff}(\varphi )&=&-\frac{1}{2}~tr~[\log ~(G_{0}^{-1} +V)
+\frac{1}{2}(\partial_{\mu}\varphi )^2 +\frac{\lambda}{4}(\varphi^2 -f_{\pi}^2
)^2 ]\nonumber\\
~&=&-\frac{1}{2}~tr~[\log ~(1+G_0 V)]+\frac{1}{2}(\partial_{\mu}\varphi )^2 
+\frac{\lambda}{4}(\varphi^2 -f_{\pi}^2
)^2 +\frac{1}{2}~tr~\log ~G_{0}.
\end{eqnarray}
The last term is an infinite constant and may be neglected. Now the
first term can be developed as

\begin{eqnarray}
\label{first}
-\frac{1}{2}~tr~[\log (1+G_{0}V)]=-\frac{1}{2}tr~ G_{0}V
+\frac{1}{4}tr~(G_{0}V)^2 -\frac{1}{6}tr~(G_{0}V)^3 +...
\end{eqnarray}

Let us explicitly evaluate the first term of the above equation in the effective action

\begin{eqnarray}
-\frac{1}{2}\int d^4 r ~tr~G_{0}V &=&-\frac{1}{2}~Tr~G_{0}V
=-\frac{1}{2}\sum_{\vec k \omega s}~ \langle \vec k \omega s |G_{0}V| \vec k \omega s
\rangle \nonumber\\
~&=& -\frac{1}{2}~Tr~V \left(\frac{1}{\beta \Omega }\sum_{\vec k ~\omega }
\frac{1}{e_{k}^2 -\omega^2 }\right).
\label{firsteq}
\end{eqnarray}
For the sake of renormalizability, in the calculations which follows we will be concerned only with the temperature independent (divergent) terms. The sum over $\omega $ yields \cite{Grad}

\begin{eqnarray}
\frac{1}{\beta }\sum_{\omega }\frac{1}{e_{k}^2 -\omega^2 }=
\int_{-i\infty }^{+i\infty }\frac{dw}{2\pi i}~ \frac{1}{e_{k}^{2}-w^2 }=
\int_{-\infty }^{+\infty }\frac{dx}{2\pi}~ \frac{1}{e_{k}^{2}+x^2 }=
\frac{1}{2e_{k}}. 
\end{eqnarray}
The sum over ${k}$ will diverge and it is done looking at the limit $\Omega \to \infty$

\begin{equation}
\sum_{k}= \lim_{\Lambda\rightarrow \infty}
\frac{\Omega}{(2\pi )^3 }4\pi \int_{0}^{\Lambda} k^2 dk ,
\end{equation}
and analyzing the behavior for $\Lambda \to \infty$
\begin{eqnarray}
\frac{1}{\Omega} \sum_{\vec{k}} \frac{1}{e_k} &=& \frac{1}{(2\pi )^3 }4\pi \int_{0}^{\Lambda} \frac{k^2}{\sqrt{k^2+m^2}} dk =
\frac{1}{2\pi^2} \frac{m^2}{2} \left[\frac{sinh2x}{2}-x \right]_{\Lambda \to \infty}
\sim \nonumber\\
~&\sim & \frac{1}{2\pi^2 }~\frac{m^2 }{2}\left[\left(
\frac{\Lambda}{m}\right )^2 -
\log~\left(\frac{2 \Lambda}{m}\right) + \frac{1}{2}\right ]+O\left (
\frac{m^2}{\Lambda^2}\right ),
\end{eqnarray}
where $m=gf_{\pi}$, $\frac{\Lambda}{m}=sinhx$ and $x=\log \left(\frac{\Lambda}{m}+ \sqrt{1+\frac{\Lambda^2}{m^2}} \right)$. The coefficient in brackets is therefore
 
\begin{eqnarray}
\frac{1}{\beta \Omega }\sum_{\vec{k} \omega }~ \frac{1}{e_{k}^2
-\omega^2 }\longrightarrow_{\Lambda\to \infty}
\frac{1}{2\pi^2} \frac{m^2}{4} \left[\left(\frac{\Lambda}{m}\right)^2 -
\log \left(\frac{2\Lambda}{m}\right )  +
\frac{1}{2}\right ]+O\left(\frac{m^2 }
{\Lambda^2}\right).
\end{eqnarray}
We can therefore write

\begin{eqnarray}
-\frac{1}{2}Tr~G_{0}V= -a_1 (\Lambda )\int d^4r g(\varphi^2 -\varphi_{0}^2 ),
\end{eqnarray}
where $a_1 (\Lambda )\longrightarrow_{\Lambda\to \infty} \Lambda^2$. We could had guess this behavior, since $Tr G_0V \sim \int \frac{d^4k}{k^2} \sim \Lambda^2$.

The next term in the expansion contain $G_0^2 \sim \frac{1}{k^4 }$
so that $Tr(G_0 V)^2 $ will have a logarithmic divergence. The other
terms will be finite and therefore we need not worry about them. Let us
then proceed to the evaluation of the second term in the r.h.s. of eq.(\ref{first}):

\begin{eqnarray}
\frac{1}{4}\int d^4r tr (G_0V)^2 = \frac{1}{4}~Tr~ G_0 V G_0 V&=& \frac{1}{4}~Tr~G_0^2 V^2 +\frac{1}{2} 
~Tr~[G_0,V]^2 \nonumber\\
&=&Tr~( G_0^2 V^2)+ \frac{1}{2} Tr~G_0^2 ~[V,G_0^{-1}]~G_{0}^2 ~[V,G_{0}^{-1}],
\end{eqnarray}

\begin{eqnarray}
Tr~G_0 V G_0 V&=& Tr~ G_0^2 V^2 +\frac{1}{2} Tr~ G_0^4 ~[V,G_0^{-1}]^2
\nonumber\\
~&+&\frac{1}{4} Tr~ G_0^8 ~[[V,G_0^{-1}],G_0^{-2}]+....
\end{eqnarray}
We thus have

\begin{eqnarray}
\frac{1}{4}Tr~ G_0 V G_0 V&=&\frac{1}{4}\sum_{k\omega S}\frac{ \langle 
k\omega S| V^2
|k\omega S\rangle }{e_k^2 -\omega^2 }
+\frac{1}{2}\sum_{k\omega S}\frac{\langle k\omega
S |[V,G_0^{-1}]^2
|k\omega S\rangle }{(e_k^2 -\omega^2 )^4 }+...\nonumber\\
&=&\frac{1}{4}Tr~V^2 \left( \frac{1}{\beta \Omega }\sum_{k\omega }\frac{1}
{(e_k^2 -\omega^2)^2 }\right)+\frac{1}{8}\sum_{k\omega S}\frac{\langle
k\omega S | [V,G_0^{-1}]^2
|k\omega S\rangle }{(e_k^2 -\omega^2 )^4 }+...
\label{g0vg0v}
\end{eqnarray}
The second term in eq.(\ref{g0vg0v}) will be finite as can be easily checked by
power counting. For the first term we have

\begin{eqnarray}
\frac{1}{\beta }\sum_{\omega}~\frac{1}
{(e_k^2 -\omega^2 )^2 }=\frac{1}{4~e_k^3 }.
\end{eqnarray}
Now \cite{Grad}

\begin{eqnarray}
\frac{1}{\Omega }\sum_k \frac{1}{e_k^3 }&=&\frac{1}{(2\pi)^3 }~4\pi
\int_0^{\Lambda} \frac{k^2 dk }{(k^2 +m^2 )^{3/2}}\nonumber\\
&\longrightarrow_{\Lambda\to \infty}&\frac{1}{2\pi^2}
\left(\log \left(\frac{2\Lambda}{m}\right)-1\right)+
O\left(\frac{m^2 }{\Lambda^2}\right). 
\end{eqnarray}
We can thus write, in analogy to eq.(\ref{firsteq})

\begin{eqnarray}
\frac{1}{4}Tr(G_0 V)^2 =a_2 (\Lambda )\int d^4r \left[g^4 
(\varphi^2 -\varphi_0^2 )^2
+g^2 (\partial_{\mu}\varphi )^2 \right ]+finite~ contributions,
\end{eqnarray}
where now $a_2 (\Lambda )\longrightarrow_{\Lambda\to \infty} 
\log~\left(\frac{\Lambda }{m} \right )$.


Having identified the divergent part of the effective Lagrangian we can
add counter-terms which will cancel the divergences and which will be of
the form of terms already present in the Lagrangian.
Consider then

\begin{eqnarray}
{\cal L}_{eff}&=& c_0 +\frac{1}{2}~tr~\log~G_0 +\frac{1}{2}
\partial_{\mu} \varphi
^2 + a_2 (\Lambda)(\partial_{\mu} \varphi )^2 g^2 +c_2 (\varphi^2 -f_{\pi}^2
) \nonumber\\
~&-& a_1 (\Lambda )g^2 (\varphi^2 -f_{\pi}^2 )+c_4 (\varphi^2 -f_{\pi}^2 )^2
+a_2 (\Lambda )g^4 (\varphi^2 -f_{\pi}^2 )^2.
\label{leff}
\end{eqnarray}
We next introduce changes in the parameters $c_0$, $c_2$ e $c_4$ on
the one hand and proceed to a redefinition of $\varphi $, $g$ and $f_{\pi}$
as follows

\vspace{0.3cm}

1) $c_0 $ is chosen so that $c_0 +\frac{1}{2} tr~\log~ G_0 =0.$

\noindent This choice is equivalent to defining the zero in energy scale and the
vacuum will have zero energy.

\vspace{0.3cm}

2) The redefinition of $\varphi $:

\begin{eqnarray}
\varphi\rightarrow \varphi_R =\frac{1}{\sqrt{1+2c_2 g^2 }}~\varphi.
\end{eqnarray}
Thus the third and fourth terms in eq.(\ref{leff}) becomes

\begin{eqnarray}
\frac{1}{2}(\partial_{\mu} \varphi )^2 +a_2 ~g^2 ~(\partial_{\mu} \varphi )^2 
\rightarrow \frac{1}{2}(\partial_{\mu}\varphi )^2.
\end{eqnarray}
This is the field renormalization. Also, accordingly 

\begin{eqnarray}
f_{\pi}\rightarrow f_{\pi}^R =\frac{1}{\sqrt{1+2c_2 g^2 }}f_{\pi},
\end{eqnarray}
and

\begin{eqnarray}
\varphi^2 -f_{\pi}^2 \rightarrow \frac{1}{1+2a_2 g^2 }(\varphi^2 -f_{\pi}^2 ).
\end{eqnarray}

\vspace{0.3cm}

3) Also, in order to maintain the form of $G_0^{-1} =-\partial_{\mu}^2
+g^2 f_{\pi}^2 $ we need to redefine the coupling constant

\begin{eqnarray}
g\rightarrow g_R =\sqrt{1+2a_2 g^2 }~g.
\end{eqnarray}

\vspace{0.3cm}

4) The coefficient $c_2 $  is chosen so that

\begin{eqnarray}
c_2 - a_1 ~g_R^{2} =0.
\end{eqnarray}

\vspace{0.3cm}

5) The coefficient $c_4 $ is chosen so that, for a given $\lambda$ , one has

\begin{eqnarray}
(c_4 +a_2 ~g_R^{4} )(\varphi_R^2 -f_{\pi R}^2 )^2 =\frac{\lambda}{4} 
(\varphi_R^2-{f_{\pi R}^2} )^2.
\end{eqnarray}

The renormalized lagrangean is then given by

\begin{eqnarray}
\label{L}
{\cal L}_{eff}(\varphi )&=&-\frac{1}{2}~tr~\log~(1+G_0 V)+\frac{1}{2}~tr~ G_0 V
-\frac{1}{4}~tr~G_0^2 V^2 \nonumber\\
~&+& \frac{1}{2} (\partial_{\mu} \varphi_{R} )^2 
+\frac{\lambda}{4}(\varphi_R^2 -{f_{\pi R}^2} )^2.
\end{eqnarray}

Summarizing, we have shown that the `` bosonized '' Lagrangian was rendered ultraviolet finite with no necessity of introducing new parameters to the theory \cite{Landshoff}. The partition function $Z$ remains thus finite and the counterterms needed are the same as in the usual zero temperature renormalization schemes. The effective meson Lagrangian (\ref{L}) gives an effective action which is a finite functional of $\varphi$. The physical 
`` memory '' of the fermion-meson interaction is totally contained in the finite parts of $ -\frac{1}{2}tr~\log~(1+G_0 V)$ that are $-\frac{1}{6}tr~(G_{0}V)^3 + \frac{1}{8}tr~(G_{0}V)^4 +...$. These terms will enter in the potential energy density part of $-\frac{T}{V} \ln Z$. From $Z$ all important information concerning the thermodynamic macroscopic properties of the effective system in equilibrium can be obtained.

\vspace{0.5cm}

{\bf Acknowledgements}\\
This work was partly supported by CNPq. H.C. thanks A.L. Mota and M. Hott for useful discussions and for a critical reading of the manuscript.




\begin{references} 
\let\ul=\underbar
\def\AP#1{{Ann. Phys. }{\bf #1}}
\def\PRP#1{{Phys. Rep. }{\bf #1}}
\def\APP#1{{Act. Phys. Pol. }{\bf #1}}
\def\PTP#1{{Prog. Theor. Phys. }{\bf #1}}
\def\PR#1{{Phys. Rev. }{\bf #1}}
\def\PRD#1{{Phys. Rev. }{\bf D #1}}
\def\PRC#1{{Phys. Rev. }{\bf C #1}}
\def\PRL#1{{Phys. Rev. Lett. }{\bf #1}}
\def\PL#1{{Phys. Lett. }{\bf #1}}
\def\RMP#1{{Rev. Mod. Phys.}{\bf #1}}
\def\NP#1{{Nucl. Phys. }{\bf #1}}
\def\ZP#1{{Z. Phys. }{\bf #1}}
\def\NC#1{{Nuovo Cimento }{\bf #1}}
\def\SJNP#1{{Sov. J. Nucl. Phys. }{\bf #1}} 

\bibitem{Dolan} L. Dolan and R. Jackiw, Phys. Rev. D {\bf 9}, 3320 (1974).
\bibitem{Weinberg} S. Weinberg, \PRD{\bf 9}, 3357 (1974).
\bibitem{Bernard} C.W. Bernard, \PRD {\bf 9}, 3312 (1974).
\bibitem{Cornwall} J.M. Cornwall, R. Jackiw, and E. Tomboulis,
\PRD{10}, 2428 (1974).
\bibitem{Baym} G. Baym and G. Grinstein, \PRD{15}, 2897 (1977).
\bibitem{Linde} A. Linde, Rep. Prog. Phys. \NC{42}, 389 (1979). 
\bibitem{Kapusta} J. Kapusta, {\it Finite-Temperature Field Theory}
(Cambridge University Press, Cambridge, 1989).
\bibitem{Braaten0} E. Braaten and R.D. Pisarski, \NP{B337}, 569 (1990); E. Braaten and R.D. Pisarski, \NP{B337}, 369 (1990); E. Braaten and R.D. Pisarski, \PRL{64}, 1338 (1990).
\bibitem{Bellac} Michel Le Bellac, {\it Thermal Field Theory}
(Cambridge University Press, Cambridge, 1996).
\bibitem{Das} A. Das, {\it Finite-Temperature Field Theory} (World Scientific, NY, 1997).
\bibitem{Caldas} H.C.G. Caldas, D.H.T. Franco, A.L. Mota, F.A. Oliveira and M.C. Nemes, \NP{A 617}, 464 (1997).
\bibitem{Caldas2} H.C.G. Caldas, A.L. Mota, and M.C. Nemes, \PRD{63}, 56011 (2001).
\bibitem{Gell-Mann} M. Gell-Mann, and M. Levy, \NC{16}, 705 (1960).
\bibitem{Lee} B. Lee, {\it Chiral Dynamics} (Gordon and Breach, 1970).
\bibitem{Franco} D.H.T. Franco, H.C.G. Caldas, A.L. Mota and M.C. Nemes, Mod. Phys. Lett. A617, 464 (1997) and references therein.
\bibitem{Ram-Mohan} L.R. Ram Mohan, \PRD{14}, 2670 (1976).
\bibitem{Caldas3} H.C.G. Caldas, to appear in Phys. Rev. D, hep-th/0111194.
\bibitem{Dirac} C. Itzykson, J.B. Zuber {\it Quantum Field Theory} (MacGraw-Hill, 1985).
\bibitem{Grad} I.S. Gradshteyn, I.M. Ryzhik {\it Table of Integrals, Series, and Products} (Academic Press, 6th edition, 2000).
\bibitem{Landshoff} D. Bodeker, P.V. Landshoff, O. Nachtmann and A. Rebhan, 
\NP{B539}, 233 (1999). 


\end{references}
\end{document}